\documentclass[10pt]{article}
\usepackage{extsizes}
\usepackage[super,sort&compress,comma]{natbib} 
\usepackage[version=3]{mhchem}
\usepackage[left=1.5cm, right=1.5cm, top=1.785cm, bottom=2.0cm]{geometry}
\usepackage{lscape}
\usepackage{balance}
\usepackage{times,mathptmx}
\usepackage{sectsty}
\usepackage{graphicx} 
\usepackage{lastpage}
\usepackage{float}
\usepackage[english]{babel}
\usepackage{array}
\usepackage{charter}
\usepackage{setspace}

\usepackage{epstopdf}

\newcommand{\nanairo}{Na$_2$IrO$_3$}
\newcommand{\nairo}{NaIrO$_3$}
\newcommand{\liiro}{Li$_2$IrO$_3$}

\newcommand{\srcairo}{Sr$_3$CaIr$_2$O$_9$}
\newcommand{\srcaruo}{Sr$_3$CaRu$_2$O$_9$}
\newcommand{\via}{{\it via}}
\newcommand{\angstrom}{\mbox{\normalfont\angstrom}}

\begin{document}

\title{New Honeycomb Iridium (V) Oxides: \nairo~and \srcairo.}

\author{
	David C. Wallace$^{1,2}$ and Tyrel M. McQueen$^{1,2,3,*}$\\ \\
	$^{1}$Department of Chemistry, The Johns Hopkins University, Baltimore, MD 21218, USA\\
	$^{2}$Institute for Quantum Matter, Department of Physics and Astronomy,\\The Johns Hopkins University, Baltimore, MD 21218, USA\\
	$^{3}$Department of Materials Science and Engineering, \\The Johns Hopkins University, Baltimore, MD 21218, USA\\
	*{\it Email:} mcqueen@jhu.edu
	}
\maketitle
\begin{abstract}
\begin{doublespace}
We report the structures and physical properties of two new iridates, \nairo~and \srcairo, both of which contain continuous two-dimensional honeycomb connectivity. \nairo~is produced by room temperature oxidative deintercalation of sodium from \nanairo, and contains edge-sharing IrO$_6$ octahedra that form a planar honeycomb lattice. \srcairo, produced \via~conventional solid-state synthesis, hosts a buckled honeycomb lattice with novel corner-sharing connectivity between IrO$_6$ octahedra. Both of these new compounds are comprised of Ir$^{5+}$ (5d$^{\rm 4}$) and exhibit negligible magnetic susceptibility. They are thus platforms to investigate the origin of the nonmagnetic behavior exhibited by Ir$^{5+}$ oxides, and provide the first examples of a {\it J}~=~0 state on a honeycomb lattice.
\end{doublespace}
\end{abstract}


\onecolumn
\begin{doublespace}
\section{Introduction}
Research into 5d transition metal compounds has blossomed in recent years due to the potential for new electronic and magnetic phenomena to emerge from the interplay between crystal field effects, electronic correlations (Hubbard U) and strong spin-orbit coupling (SOC). Iridium is an appealing element in which to study these physics due to its tendency to adopt a 4+ oxidation state, where the ground electronic configuration 5d$^5$ is predicted to yield a {\it J}~=~$\frac{1}{2}$ electronic state in the presence of strong SOC \cite{MottStrongSOC}. Further, iridates are known to crystallize in numerous structure types, each with unique Ir-Ir connectivity, and this fact has made them an active area of study\cite{Yang-Kim,Na2IrO3Synthesis,Liu,BPhelan}. In aggregate, recent literature on the magnetic and electronic properties of iridates are unified by a desire to probe the influence of strong SOC and U on electronic behavior\cite{BalentsSOCReview}. Iridium--based honeycombs, in particular the {\it A}$_2$IrO$_3$ ({\it A}~=~Li/Na) family, have fallen under intense scrutiny following predictions that they might host a spin-liquid ground state\cite{Na2IrO3Prediction,XtalfieldandcorrelationA2IrO3,A2IrO3KitaevHeisenberg,Kimchi,Andrade,Rau,Lei,Katukuri}. Efforts to produce new iridium oxide honeycomb lattices have intensified\cite{ThinFilmHoneycomb,HarmonicHoneycomb, BaroudiCava}, but 4d and 5d materials containing continuous honeycomb connectivity remain scarce.

Here we report the synthesis and characterization of two new honeycomb iridates based on Ir$^{\rm 5+}$ (5d$^{\rm 4}$).
The first, \nairo~is produced via~{\it chimie douce} oxidative deintercalation of sodium from \nanairo, which preserves the planar honeycomb network of edge-sharing IrO$_6$ octahedra. The second, \srcairo, is a 2:1 ordered perovskite, which shares its structure with Sr$_3$CaRu$_2$O$_9$\cite{Poeppelmeier}, and consists of layers of IrO$_6$ octahedra forming a buckled honeycomb lattice. Magnetic susceptibility data collected on both compounds demonstrate weak temperature-independent magnetism, and resistivity measurements show semiconducting behavior consistent with hopping conductivity. These compounds are thus ideal to investigate the interplay between non-cubic crystal fields and SOC. By comparison with related 4d and 5d honeycomb compounds with partially filled t$_{\rm 2g}$ subshells, we find that d electron count correlates strongly with the magnetic behavior, and present a thorough structural analysis of \srcairo~and \srcaruo~that ultimately suggests the correlation is driven by SOC.
\section{Results}
\subsection{Syntheses and Structures}

\nairo~was prepared \via~the oxidative deintercalation reaction: \begin{equation}\rm Na_2IrO_3+\frac{x}{2}~Br_2~\underset{Acetonitrile}{\longrightarrow}~(1-x)~Na_{2}IrO_3+x~NaIrO_3+x~NaBr.\end{equation} The 1:1 relationship between oxidant consumption and Na$^+$ deintercalation of equation 1 was confirmed by laboratory X-ray powder diffraction (XRPD) data collected on powders after vacuum evaporation of the solvent, which showed crystalline NaBr as the only side product. In contrast to the precursor material \nanairo~which degrades rapidly in laboratory air\cite{Krizan}, \nairo~is stable under ambient conditions, as XRPD data showed no evidence of decomposition after several months of air exposure.  However, \nairo~degrades rapidly upon heating above 200$^{\circ}$C in air, as evidenced by the broadening of diffraction peaks and appearance of new reflections. Attempts at producing intermediate stoichiometries (i.e. Na$_{2-x}$IrO$_3$) were unsuccessful, and instead produced two-phase samples containing a mixture of \nanairo~and \nairo. Further, no phase width was observed in \nairo, as reactions performed with excess bromine produced only \nairo~with no observable difference in lattice parameters or physical properties. 

\begin{figure}
\centering
\includegraphics[width=3.5in]{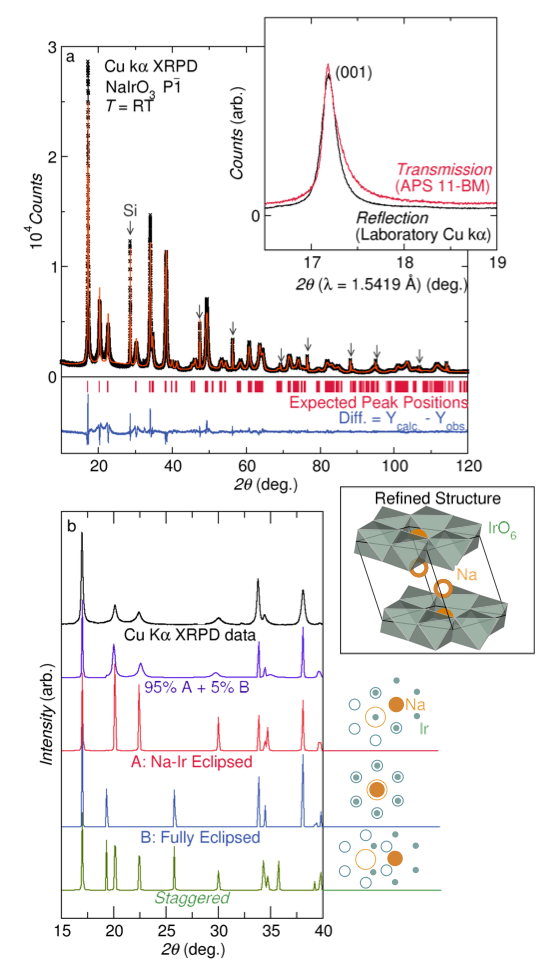}
\caption{{\bf a)} Laboratory XRPD data (black) collected on \nairo~along with Rietveld refinement of the structure in P$\bar{1}$ to the data (red) and calculated difference between the model and the data (blue) are shown in the main frame. Stacking faults between honeycomb layers complicate the refinement, as can be seen from the missed intensities and peak shape mismatches between 2$\theta$~=~15$^{\circ}$ to 35$^{\circ}$. In the inset, the observed profile of the (001) peak, as observed by laboratory XRPD data collected in reflection geometry, is compared to the profile observed from synchrotron XRPD data collected on APS 11-BM in transmission geometry. {\bf b)} Laboratory XRPD data (black, top) along with simulations for three distinct stacking arrangements of adjacent NaIr$_2$ honeycomb layers (red, blue and green) are shown. A fourth simulation (purple) shows the pattern that results from having a 95\% probability of an Na--Ir eclipsed stacking arrangement with a 5\% probability of a fully eclipsed stacking fault, which provides a good description of the observed diffraction pattern. In the upper right-hand corner the structure of \nairo~is shown, in which IrO$_6$ octahedra (green) share edges to form honeycomb layers with Na cations (orange) occupying the honeycomb "holes" and partially occupying the spaces sandwiched between Na and Ir sites in the honeycomb layers.}
\label{NaIrO3_rietveld}
\end{figure}

The structure of \nairo~was solved in space group $P\bar{1}$ \via~Rietveld refinement to laboratory XRPD data (Fig.~\ref{NaIrO3_rietveld}(a)). Final structural parameters are available in the SI. Solution of the structure in higher symmetry cells was prevented by the high degree of inter-plane stacking disorder, which arises due to the fact that there are several possible translations that can take place in the honeycomb plane from one layer to another, each of which generates a different stacking pattern at minimal energetic cost. Stacking disorder is a well known structural perturbation in many layered compounds, and evidence of this disorder can be plainly seen in the poorly-fit peak broadening and incorrect intensities present at low diffraction angles. Further evidence of this structural disorder is provided by synchrotron XRPD data (Fig.~\ref{NaIrO3_rietveld} (a, inset)), which show highly asymmetric $\langle$0 0 l$\rangle$ peaks--this may occur as a consequence of decomposition in the intense synchrotron beam. Using DIFFaX\cite{diffax}, we modeled the diffraction patterns for three distinct stacking variants. When viewed from perpendicular to the honeycomb plane, these stacking variants can be described based on which atoms in the honeycomb plane eclipse (sit directly above ) the Na ion in the adjacent plane. The three primary stacking possibilities are: 1) Na-Ir eclipsed; 2) Na-Na fully eclipsed, which produces infinite Na channels; 3)~staggered, in which none of the atoms in the honeycomb plane eclipse those in the adjacent plane, which is the pattern adopted by the parent material \nanairo. The simulated diffraction patterns for these stacking arrangements, along with XRPD data collected on a sample of \nanairo~are shown in figure Fig.~\ref{NaIrO3_rietveld}(b). The observed XRPD pattern closely resembles the simulation obtained for the Na-Ir eclipsed variant, but with noticeable broadening and attenuation of several peaks. A fourth pattern was simulated based on a model consisting of 95\% Na-Ir eclipsed stacking with a 5\% probability of a fully-eclipsed stacking fault. This model produces a good qualitative agreement with the raw XRPD data, suggesting that the fully eclipsed stacking variant may only be slightly less energetically favorable than the Na-Ir eclipsed variant. The presence of the fully eclipsed stacking fault suggests that further synthetic work may yield control over stacking order in this compound. For an in-depth discussion on stacking disorder in honeycomb iridates, see ref.~\cite{Wallace}. This is the second known structure with the formula \nairo-a post-perovksite structure with the same formula is obtained when synthesized under high-pressure and temperature \cite{NaIrO3_postPerov}. Neither structural polymorph of \nairo~can be synthesized using conventional solid-state techniques, a fact that underscores the novelty and relative instability of the Ir$^{5+}$ oxidation state.

\begin{figure}
\centering
\includegraphics[width=7in]{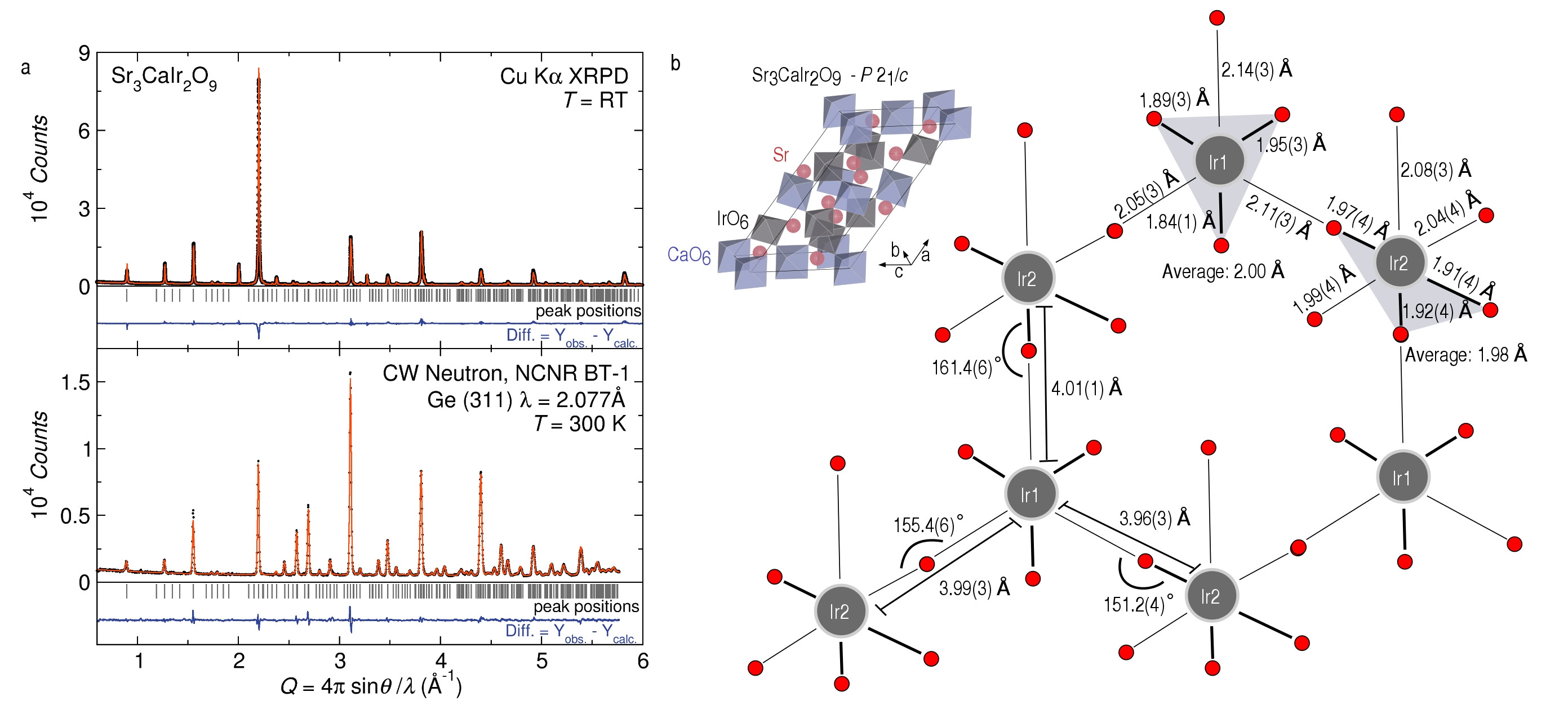}
\caption{{\bf a)} Rietveld refinement of the structure of \srcairo~to laboratory XRPD data (top) and NCNR BT-1 NPD data (bottom) are shown. Data points are displayed in black, the fit in orange, and the difference between the data and fit, blue. Expected peak positions for the monoclinic cell are displayed as gray tick marks below the data in each pane. {\bf b)} The unit cell of \srcairo~is displayed along with a cutout of the honeycomb network formed by corner-sharing IrO$_6$ octahedra in the a-b plane. Ir--Ir distances and Ir--O--Ir bond angles are shown in the bottom left of the honeycomb structure, and Ir--O bond lengths for each of the two iridium sites are shown in the upper right hand corner. Despite the uniformity of the Ir--Ir distances in the honeycomb lattice, the IrO$_6$ octahedra exhibit significant asymmetric distortions, and these distortions are illustrated by gray triangles in the plane formed by the three most closely bonded oxygen atoms of each octahedron. These distortions are inconsistent with a Jahn-Teller effect, and instead are likely driven by strong SOC of the Ir$^{\rm 5+}$ ion as described in the text.}
\label{srcairo_structure}
\end{figure}

The structure of \srcairo~was solved in space group {\it P}2$_{\rm 1}$/{\it c} (14) \via~Rietveld refinement to neutron powder diffraction (NPD) and XRPD data (Fig.~\ref{srcairo_structure} (a)), and was found to be isostructural to \srcaruo\cite{Poeppelmeier}. Final structural parameters are available in the SI. The monoclinic structure of \srcairo, visible in Fig.~\ref{srcairo_structure} (b), can be described as a 2:1 ordered perovskite with the formula A$_3$BB'$_2$O$_9$, where IrO$_6$ and CaO$_6$ octahedra share corners to form the perovskite lattice and Sr$^{2+}$ cations occupy the 12-fold coordinate A sites. In contrast to the Ba$_3$$M$Ir$_2$O$_9$ ({\it M} = Mg, Ca, Sc, Ti, Zn, Sr, Zr, Cd, In, etc.) family of 6H-perovskite-like structures, which host face-sharing Ir$_2$O$_9$ dimers\cite{Hinatsu}, the IrO$_6$ octahedra of \srcairo~share corners to form a buckled honeycomb lattice in the b--c plane. While the structure of \srcairo~may seem odd given the apparent wealth of other Barium-based compounds with analogous stoichiometry, it is actually unsurprising that a conventional perovskite lattice is formed in this case due to the improved match in ionic radii between Sr and Ir. The same is true of \srcaruo, which also has a similar collection of Ba-based 6H-perovskite-type cousins. While many different perovskites containing iridium have been reported in addition to the widely studied Sr$_2$IrO$_4$, this seems to be a rare example of an iridium-based perovskite containing honeycomb connectivity. Furthermore, the corner-sharing bonding motif of the honeycomb lattice is the first of its kind in iridates, and thus opens a new avenue along which to search for new honeycomb materials that are free from stacking disorder and are air--stable. 

The corefinement of the \srcairo~structure to both XRPD and NPD data yielded stable oxygen positions that can be used to investigate the honeycomb connectivity in detail (Fig.~\ref{srcairo_structure} (b)). As a result of octahedral tilting, the Ir--O--Ir bond angles between nearest-neighbor sites on the honeycomb lattice are significantly less than the ideal 180$^{\circ}$ expected for an undistorted perovskite. The Ir--O bond lengths are also distorted: for the first iridium site, three adjacent Ir--O bonds are significantly shorter than the overall average of 1.99 \AA, while the opposite three are significantly longer. A similar "three short, three long" pattern is also observed to a lesser degree on the second iridium site, where the three short bonds form a plane that is roughly perpendicular to the plane formed by the longer three bonds. This pattern results in significant variations in the bridging Ir--O--Ir bond lengths: the shortest Ir--Ir distance (3.96(3) \AA) is formed via Ir--O bonds of very similar lengths (2.04(4) \& 2.05(3) \AA, respectively), while the longest (d$_{\rm Ir-Ir}$ = 4.01(1) \AA) is formed via Ir--O bonds of very disproportionate lengths (2.14(3) \& 1.92(4) \AA, respectively). These non-uniform exchange interactions result in an effective dimerization between Ir sites even though the distances between adjacent iridium sites on the honeycomb lattice are remarkably similar. These observations are consistent with a non-spherical perturbation to the d-electron states of Ir 5+ , which will be discussed in detail in the next section.
\subsection{Physical Properties}

\begin{figure}
\centering
\includegraphics[width=3.5in]{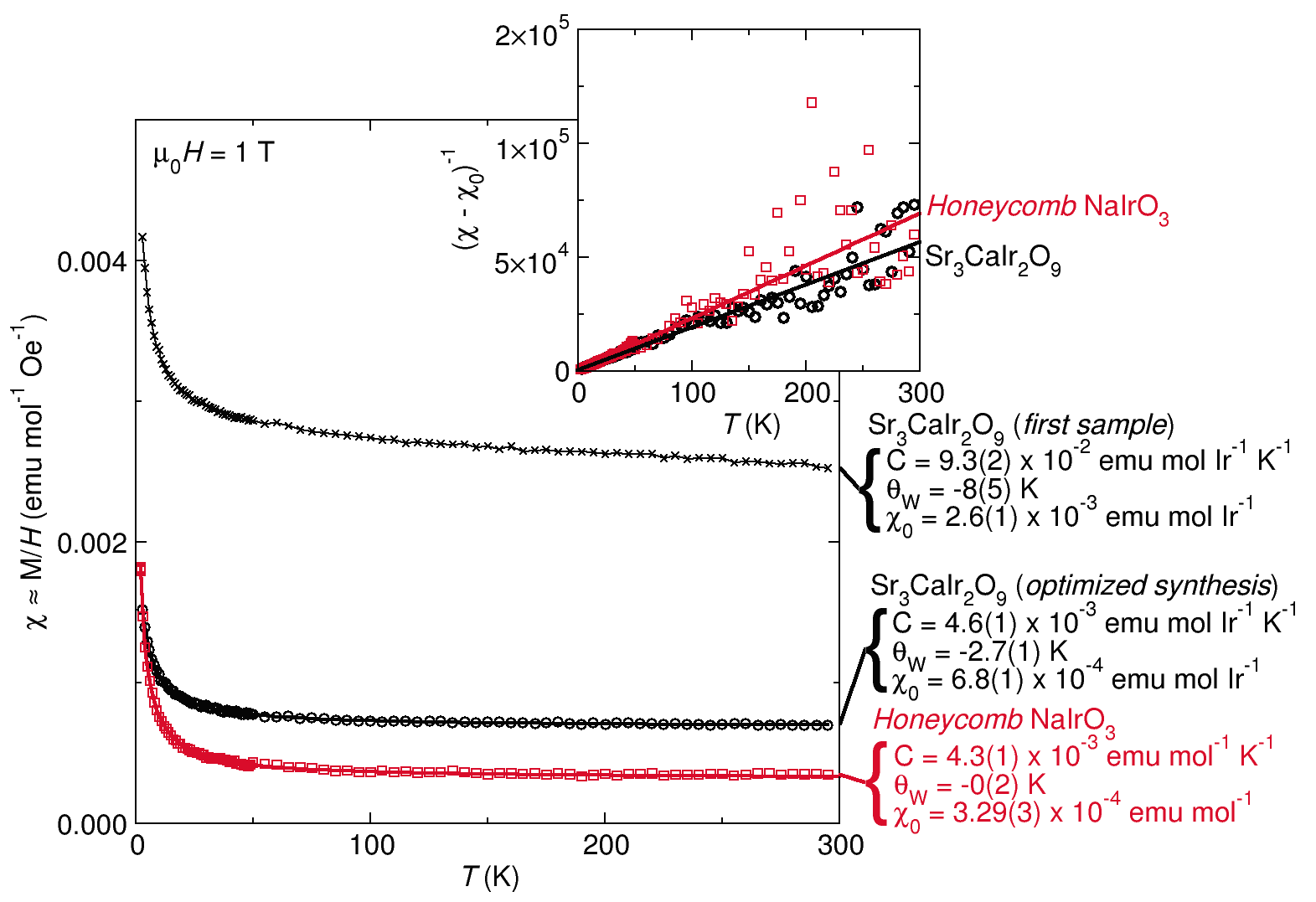}
\caption{Magnetic susceptibility data collected on \nairo~(red squares), the first sample of \srcairo~(black crosses), and a sample of \srcairo~produced \via~an optimized synthetic procedure (black circles) are shown. All magnetic data were collected on polycrystalline samples, and a correction for the temperature-independent susceptibility of the sample holder was applied to the data in order to extract real values of $\chi_0$ for each sample. Two datasets are shown for \srcairo~to illustrate that the observed sample $\chi_0$'s are due to magnetic defects, the concentration of which can be substantially reduced through correct sample preparation. Linear fits to the $\chi_0$-corrected inverse susceptibility data for \nairo~and \srcairo, visible in the inset, were used to extract Curie constants C and Weiss temperatures $\Theta_W$ for \nairo~and the optimized sample of \srcairo. Both samples exhibit weak, temperature-independent susceptibility consistent with a small concentration of magnetic defects. These data demonstrate that the Ir$^{\rm 5+}$ species in the bulk of both compounds have no unpaired electrons.}
\label{honeycomb_magnetization}
\end{figure}

Fig.~\ref{honeycomb_magnetization} shows magnetic susceptibility data collected on \nairo~and \srcairo, along with linear fits to the $\chi_0$-corrected inverse susceptibilities (inset). In contrast to \nanairo, which has a magnetic moment of 2.0(1) $\mu_B$ per Ir site and a Weiss temperature $\theta_W$~=~-159(3)~K, honeycomb \nairo~shows only a small temperature-independent susceptibility $\chi_0$~=~3.29$\times$10$^{-4}$~emu/mol~Ir in the high temperature regime and a Curie tail at low temperatures. A Curie-Weiss fit to the magnetic susceptibility data (Fig.~\ref{honeycomb_magnetization} (inset)) yields a Curie constant of C~=~4.3(1)$\times$10$^{-3}$ emu/mol Ir K, and a Weiss temperature $\theta_W$~=~0(2) K. Similar magnetic behavior is observed in \srcairo~(Fig.~\ref{honeycomb_magnetization}). \srcairo~was initially obtained as a phase-pure powder after 6 short heatings, and a Curie-Weiss fit to the magnetic susceptibility data collected on this sample (Fig.~\ref{honeycomb_magnetization} (inset)) yielded a temperature-independent susceptibility of  $\chi_0$~=~2.6(1)$\times$10$^{-3}$~emu/mol~Ir, a Curie constant C~=~9.3(1)$\times$10$^{-3}$ emu/mol Ir K, and a Weiss temperature $\theta_W$~=~-8(5)~K. Optimization of the stoichiometry and heating schedule for \srcairo~(See E.S.I.) yielded a sample whose diffraction peaks were significantly sharper and more intense, consistent with an improvement in crystallinity and homogeneity. This improvement resulted in a substantial reduction of C, $\Theta_W$ and $\chi_0$  (4.6(1)$\times$10$^{-3}$ emu/mol Ir K, -8(5)~K, and 3.29$\times$10$^{-4}$~emu/mol~Ir, respectively). In all samples, the small observed magnetic moments are likely due to dilute magnetic impurities or orphan spins--roughly 1~\% of free {\it S}~=~$\frac{1}{2}$ spins in the bulk could account for the observed susceptibility. Furthermore, the small Weiss temperatures observed from all datasets indicate that the magnetic electrons are non-interacting, which lends further weight to the argument for magnetic defects and impurities rather than intrinsic magnetism. These results are thus consistent with a {\it J}~=~0 state in the Ir$^{5+}$ metal centers of both compounds. Furthermore, the magnetic moments observed for Ir$^{\rm 5+}$ in these compounds are consistent with what is observed in many other Ir$^{\rm 5+}$ compounds (Table 1).

\begin{figure}
\centering
\includegraphics[width=3.5in]{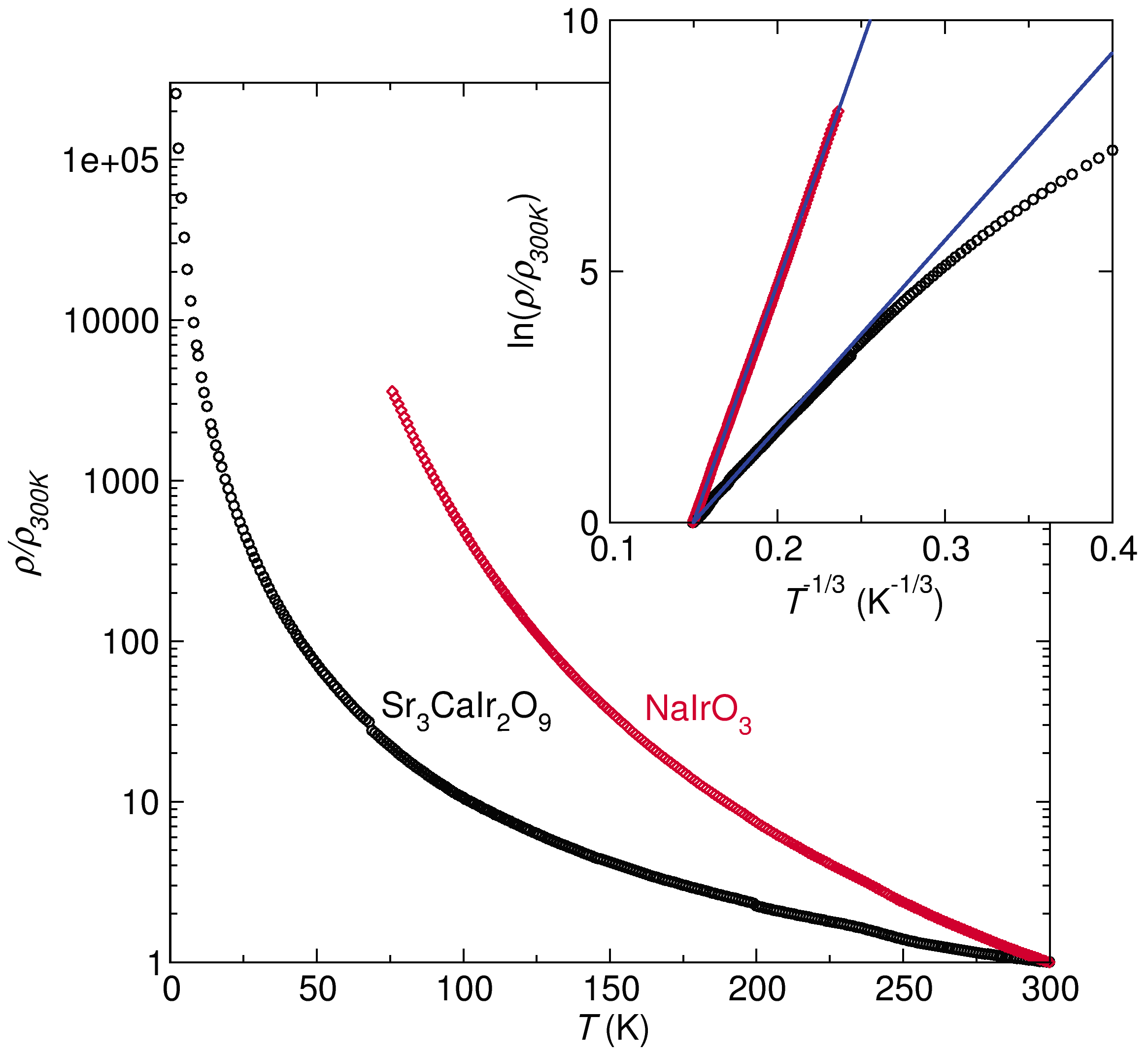}
\caption{Four-probe resistivity data collected as a function of temperature on sintered polycrystalline bars of \nairo~(red diamonds) and \srcairo~ (black circles) are shown. The increase in resistance of the pellets on cooling is a hallmark of electrically insulating materials. In the inset, the natural logarithm of the normalized resistance is plotted as a function of {\it T}$^{-\frac{1}{3}}$. Both datasets are linear in this plot in the high-temperature regime consistent with 2-dimensional variable-range hopping of impurity-localized charge carriers.}
\label{resistivity}
\end{figure}

The observed temperature-independent susceptibility is not due to delocalization of charge carriers to form a metal, as is the case in other iridates such as [Rb/K]Ir$_4$O$_8$\cite{Talanov,Schoop}. Fig.~\ref{resistivity} shows electrical resistivity data collected on polycrystalline bar of \nairo~and \srcairo. In both datasets, sample resistance increases on cooling, diverging rapidly below {\it T}$\sim$100 K, and exceeding the detection limit of the instrument below {\it T}$\sim$75 K in the case of \nairo. A simple Arrhenius-like activation barrier did not yield a good fit to the resistivity data. Instead, the behavior of both \nairo~and\srcairo~is well described by a variable-range hopping model. The conductivity is consistent with hopping in two or three dimensions: a plot of ln$\rho$ vs. $T^{\rm-\frac{1}{3}}$ (Fig. 4(inset)) or ln$\rho$ vs. $T^{\rm-\frac{1}{4}}$ (not shown) yields a linear relationship. This is similar to what is found for \nanairo~and indeed many other 4d and 5d honeycombs, and consistent with the materials structure\cite{Na2IrO3Synthesis,Luo_Li2RhO3}. 

\section{Discussion}
\begin{figure}
\centering
\includegraphics[width=3.5in]{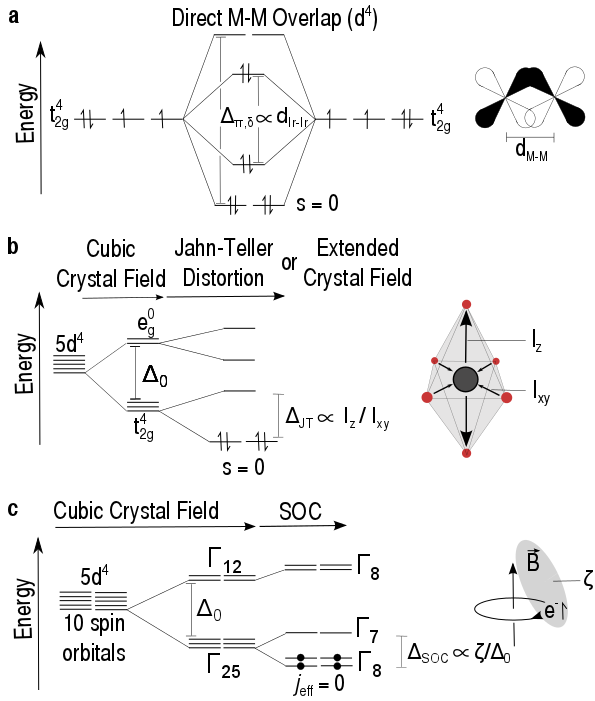}
\caption{Potential origins of {\it S}~=~0 and $J_{\rm eff}$~=~0 states in Ir$^{\rm 5+}$ iridates. {\bf a)} Direct $\pi$ and $\delta$ overlap between adjacent iridium sites yields a nonmagnetic ({\it S}~=~0) state by removing the threefold degeneracy of the isolated t$_{\rm 2g}$ manifolds. The energy of the splitting, $\Delta_{\pi,\delta}$, is proportional to the distance between adjacent iridium sites, d$_{Ir-Ir}$. {\bf b)} A local distortion of the cubic crystal field \via~Jahn-Teller distortion or interaction with cations outside the first coordination sphere (extended crystal field) splits the t$_{\rm 2g}$ manifold by an energy proportional to the axial elongation (I$_z$/I$_xy$), which also yields a {\it S}~=~0 state. {\bf c)} The ground state of the d$^{\rm 4}$ configuration in a cubic crystal field can be split by SOC to yield a {\it J}~=~0 ground state, which originates from {\it S}~=~1 and {\it L}~=~1 spin and orbital angular momenta. The upper bound to the magnitude of this splitting is proportional to the bare ion SOC constant $\xi$. relative to the crystal field $\Delta_0$.}
\label{NRGscales}
\end{figure}

Both \nairo~and \srcairo~are important new additions to a select few 4d and 5d compounds that exhibit honeycomb connectivity between MO$_6$ octahedra, tabulated in Table 2. For a given d electron count, there is little difference in observed magnetic moments between the 4d and 5d groups, despite the fact that both crystal field and SOC energy scales change significantly when moving down the periodic table. While the magnetic moments observed for the d$^5$ configurations are unsurprising given the expected magnetic moment of $\mu$$\sim$1.9$\mu_{B}$ for a {\it S}~=~$\frac{1}{2}$ system, the weak temperature-independent magnetism observed for both 4d$^4$ Ru$^{\rm 4+}$ and 5d$^4$ Ir$^{\rm 5+}$ is an unexpected result in the low-spin octahedral crystal field case, which should have 2 unpaired electrons and total spin {\it S}~=~1. Fig.~\ref{NRGscales}~shows three possible origins of a nonmagnetic state in an Ir$^{\rm 5+}$O$_6$ (Ru$^{\rm 4+}$O$_6$) octahedron. One possibility is that direct overlap between adjacent Ir sites generates new molecular orbitals, which removes the threefold degeneracy of the t$_{\rm 2g}$ manifold and results in a {\it S}~=~0 state, a scenario that has been proposed for Li$_2$RuO$_3$ based on the significant variation in Ru-Ru bondlengths\cite{Khomskii}. The second possibility is that distortions from perfect octahedral symmetry remove the degeneracy of the t$_{\rm 2g}$ manifold locally (i.e. a Jahn-Teller distortion), thus resulting in a completely filled, twofold degenerate ground state. Such distortions are thought to be highly important in the Ir$^{5+}$ perovskites [Sr/Ba]$_2$IrO$_4$, and may also be driven by extended crystal field effects (i.e. interactions beyond nearest-neighbor oxygen and iridium atoms)\cite{Bogdanov,Hozoi}. The final possibility outlined here is that SOC is the strongest perturbation to the cubic crystal field in Ir$^{\rm 5+}$, which is consistent with electronic structure calculations performed by Phelan {\it et al.} on Sr$_x$La$_{11-x}$Ir$_4$O$_{24}$\cite{BPhelan}. While the computational route invariably leads to the conclusion that SOC produces a nonmagnetic state in Ir$^{\rm 5+}$, one can also reach this conclusion on paper using group theory. To understand how SOC produces a {\it J}~=~0 state in this case, one must generate appropriate term symbols for electronic states in the presence of SOC by referencing the double group for O$_h$ symmetry. In addition to the five irreducible representations of standard character table for group O$_h$, the double group has six new irreducible representations ($\rm\Gamma_6^{+/-}$, $\rm\Gamma_7^{+/-}$, and $\rm\Gamma_8^{+/-}$)\cite{DresselhausBook}. By referencing to the O$_h$ double group, one finds that SOC splits the $\rm\Gamma_{25}^+$ (t$_{\rm 2g}$) manifold into two sets of spin orbitals: a twofold degenerate set with $\rm\Gamma_7^+$ symmetry and a fourfold degenerate set with $\Gamma_8$ symmetry. The $\rm\Gamma_{12}^+$ (e$_{\rm g}$) manifold is not split in the presence of SOC, but does acquire a new irreducible representation $\Gamma_8^+$. Because orbitals from the original $\rm\Gamma_{25}^+$ and $\rm\Gamma_{12}^+$ manifolds share the same irreducible representations, they interact to form low- and high-energy pairs, akin to bonding and antibonding orbitals in a conventional MO diagram. The net result for the d$^4$ case is a completely filled $\rm\Gamma_8^+$ manifold, which gives rise to the {\it J}~=~0 state. 

\begin{figure}
\centering
\includegraphics[width=3.5in]{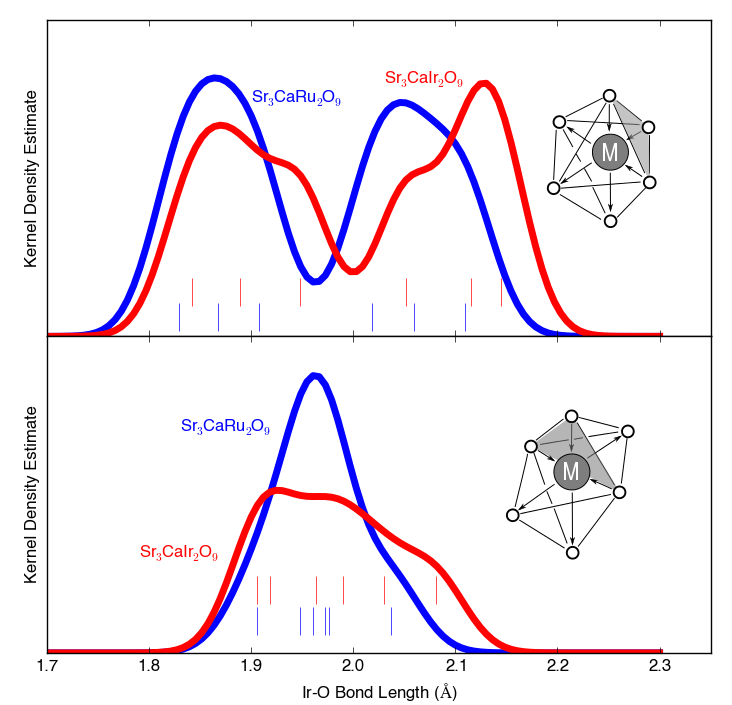}
\caption{Visual comparison of M--O (M = Ir, Ru) bond lengths in \srcairo~and\srcaruo. Gaussian kernel density estimates (KDE) computed on the Ir--O bond lengths present in \srcairo~(red) and \srcaruo~(blue) provide a visual comparison of the octahedral distortions present in both compounds. The top and bottom panes contain data for the 1st and 2nd metal sites in each structure, respectively; in each pane, the solid lines represent KDE's and the small tick marks represent the actual M--O bond lengths. A distinct "3--short, 3--long" bond length pattern is observed in both compounds. Model octahedra in each pane show the shape of the distortion in both cases, with the plane formed by the closely bonded oxygen atoms highlighted in gray.}
\label{KDE}
\end{figure}

While the first scenario (direct metal-metal bonding) is well supported for the case of Li$_2$RuO$_3$, it is not a likely explanation for our Ir$^{\rm 5+}$ compounds, as \nairo~and \srcairo~show similar magnetic properties despite having wildly different Ir-Ir connectivity and internuclear distances. Further information can be obtained by examining the Ir--O bond lengths and octahedral distortions, as they provide direct signatures of orbital degeneracy. Fig.~\ref{KDE} compares the Ir--O bond lengths present in \srcairo~and \srcaruo~using Gaussian kernel density estimates and illustrations to provide a visual understanding of the distortions present in these octahedra. Both compounds exhibit similar "three short, three long" distortions of the two octahedral metal sites, as discussed in section 2.1. These distortions are distinctly asymmetric, and thus inconsistent with a Jahn--Teller effect. The asymmetry is also likely not due to extended crystal field effects, as both Ir sites have similar extended coordination spheres (proximity to Ca$^{\rm 2+}$ and Sr$^{\rm 2+}$ ions). Rather, the observed asymmetry is likely a direct consequence of SOC, which, due to the mixing of bare-ion orbital identities, gives rise to asymmetric (i.e. direction-dependent) exchange. The fact that the octahedral distortions are amplified in \srcairo~compared to \srcaruo~provides further support for the influence of SOC. Future spectroscopic experiments can directly probe for the excited states predicted by this model, and the optical signatures should also respond to application of a magnetic field, thus yielding a powerful method of investigating SOC--driven physics in these compounds. 


\section{Conclusions}
Two new insulating honeycomb iridates, \nairo~and \srcairo, have been synthesized and characterized \via~diffraction experiments, magnetometry, and resistivity. In both compounds, iridium exists in a 5+ oxidation state, yielding a 5d$^4$ electronic configuration. Magnetization measurements demonstrate that both compounds exhibit negligible magnetic susceptibility, implying that both of these compounds are in close proximity to either a {\it S}~=~0 or a {\it J}~=~0 magnetic state. Structural studies performed on \srcairo~suggest SOC is the dominant energy scale in determining the ground magnetic state of Ir$^{\rm 5+}$ compounds. Further spectroscopic experiments on these compounds will shed light on the nature of the magnetism observed in iridates.

\section{Acknowledgements}
DCW and TMM gratefully acknowledge support from the David and Lucile Packard Foundation and the American Chemical Society Petroleum Research Fund. The Institute for Quantum Matter is supported by the U.S. Department of Energy, Office of Basic Energy Sciences, Division of Material Sciences and Engineering under Grant No. DE-FG02-08ER46544. R. DCW and TMM are grateful to the NIST Center for Neutron Research for the neutron diffraction beam time obtained on BT-1.

\end{doublespace}


\newpage

\begin{table*}
\begin{center}
\begin{tabular}{cc}
  Compound~$^{\rm ref.}$ & $\mu_{eff}$ ($\mu_{\rm B}$/Ir) \\ 
  \hline
  \hline
  \srcairo~({\it \bf this work})  & 0.18 \\ 
  Honeycomb \nairo~({\it \bf this work}) & 0.18 \\
  $\rm Sr_5La_6Ir_4O_{24}$\cite{BPhelan} & 0.3 \\ 
  Post-perovskite NaIrO$_3$\cite{NaIrO3_postPerov} & 0.28 \\ 
  Ba$_3$ZnIr$_2$O$_9$\cite{Nag} & 0.2 \\ 
  Sr$_2$YIrO$_6$\cite{Cao} & 0.91 \\ 
\end{tabular}
\caption{Comparison of magnetic moments of \nairo~and \srcairo~with several previously reported Ir$^{\rm 5+}$ compounds. With the exception of Sr$_2$YIrO$_6$, all compounds exhibit small magnetic moments consistent with a {\it J}~=~0 state in the bulk Ir$^{\rm 5+}$ species, where the measured susceptibility is due to a small magnetic impurity or contribution from polycrystalline edge states.}
\end{center}
\label{5d4_magnetism}
\end{table*}

\begin{landscape}
\begin{table*}
\centering
\begin{tabular}{ccccc}
Type & Compound~$^{\rm ref.}$ & Connectivity & Electronic configuration & Magnetic Properties \\
\hline
\hline
 {\it Planar} & SrRu$_2$O$_6$\cite{honeycombSrRu2O6,Hiley} & edge-sharing & 4d$^3$ &  AFM,  $\mu_{eff}$~=~1.43(1) $\mu_B$ \\
 & Ag$_3$LiRu$_2$O$_6$\cite{Kimber_Ag3LiRu2O6} & edge-sharing & 4d$^4$ & Weak, temperature-independent  \\
 & (Li/Na)$_2$RuO$_3$\cite{Wang_LiNaRuO3} & edge-sharing & 4d$^4$ & Weak, temperature-independent \\
 & \nairo~({\it \bf this work}) & edge-sharing & 5d$^4$ & Weak, temperature-independent \\
 & $\alpha$--RuCl$_3$\cite{honeycombRuCl3} & edge-sharing & 4d$^5$ & AFM, $\mu_{eff}$~=~2.2 $\mu_B$\\
 & Li$_2$RhO$_3$\cite{Luo_Li2RhO3} & edge-sharing & 4d$^5$ & Spin-glassy AFM , $\mu_{eff}$~=~2.0 $\mu_B$\\
 & \liiro\cite{Li2IrO3Synthesis} & edge-sharing & 5d$^5$ & AFM, $\mu_{eff}$~=~2.03 $\mu_B$\\
 & \nanairo\cite{Na2IrO3Synthesis} & edge-sharing & 5d$^5$ & AFM, $\mu_{eff}$~=~2.02 $\mu_B$\\
 & Li$_2$PtO$_3$\cite{Woodward_Li2PtO3} & edge-sharing & 5d$^6$ & Unreported \\
 \hline
{\it Harmonic} & $^\mathcal{H}\langle 1 \rangle$-Li$_2$IrO$_3$\cite{HarmonicHoneycomb} & edge-sharing & 5d$^5$ & Anisotropic, $\mu_{eff}$~=~2.03 $\mu_B$ \\
 \hline
{\it Buckled}  & \srcaruo\cite{Poeppelmeier} & corner-sharing & 4d$^3$ & Unreported \\
 & \srcairo~({\it \bf this work}) & corner-sharing & 5d$^4$ &  Weak, temperature-independent \\
 & {\it thin film} Ca$_{0.5}$Sr$_{0.5}$IrO$_3$\cite{ThinFilmHoneycomb} & corner-sharing & 5d$^5$ & Unreported \\
 \end{tabular}
 \label{ComparisonTable}
 \caption{Comparison of known 4d and 5d honeycomb compounds organized by connectivity and primary metal ion. Basic magnetic properties are listed for reference.}
\end{table*}
\end{landscape}

\end{document}